\documentclass[parindent=0pt]{article}
\usepackage{booktabs}
\usepackage{makecell}
\usepackage{subcaption}

\usepackage[utf8]{inputenc}     
\usepackage[english]{babel}     
\usepackage[a4paper, left=1in, right=1in, top=1.25in, bottom=1.25in]{geometry}
\usepackage{graphicx}           

\usepackage{amsmath,amssymb}    
\usepackage{amsthm}             
\usepackage{mathtools}          
\usepackage{enumitem}           
\usepackage{listings}           
\usepackage{todonotes}          
\usepackage{hyperref}           
\usepackage{float}
\usepackage{amsthm}
\usepackage{amsmath}
\usepackage{amscd}
\usepackage{amssymb}
\usepackage{mathrsfs}
\usepackage{pdfpages}
\usepackage{setspace}
\usepackage{physics}
\usepackage{tcolorbox}

\usepackage{xcolor}
\usepackage{tikz}
\usepackage[toc,page,title,titletoc,header]{appendix}
 \usepackage{url}
 \usepackage{babel} 
\usepackage{csquotes}
\usepackage[f]{esvect}
\usepackage{blindtext}

\newcommand{\R}{\mathbb{R}}

\newcommand{\N}{\mathbb{N}}

\newcommand{\C}{\mathbb{C}}

\newcommand{\supp}{\mbox{supp}}

\newcommand{\g}{\mathfrak{g}}

\let\phi=\varphi
\let\epsilon=\varepsilon

\theoremstyle{definition}

\newtheorem*{definition*}{\textit{Definition}}
\newtheorem*{proposition*}{\textit{Proposition}}
\newtheorem*{conjecture*}{\textit{Conjecture}}
\newtheorem{result}{Result}

\newcommand{\adots}{\mathinner{\mkern1mu \raise1pt
\hbox{.} \mkern2mu \raise4pt \hbox{.} \mkern2mu \raise7pt
\vbox{\kern7pt \hbox{.}} \mkern1mu}}
\def\[{[\![}
\def\]{]\!]}

\newlist{regimes}{enumerate}{1}
\setlist[regimes]{label={\textbullet\ Regime (\arabic*):}, leftmargin=*}
\usepackage{authblk} 

\title{%
   Finite-rank multiplicative perturbations of rotationally invariant non-Hermitian random matrices
}

  \author[1,2]{Pierre Bousseyroux\thanks{Email: pierre.bousseyroux@polytechnique.edu}}
\author[3]{Marc Potters}

\affil[1]{Econophysics Lab, Institut Louis Bachelier, 28 Pl. de la Bourse, Palais Brongniart, 75002 Paris, France}
\affil[2]{LadHyX, UMR CNRS 7646, Ecole Polytechnique, Institut Polytechnique de Paris, 91128 Palaiseau, France}
\affil[3]{Capital Fund Management, Paris, France}

\newenvironment{remarks}{
  \par\vspace{1ex}
  \noindent\textbf{Remarks.}\begin{itemize}\setlength\itemsep{0pt}}
  {\end{itemize}\par\vspace{1ex}}

\makeatletter
\renewcommand{\@fnsymbol}[1]{%
  \ifcase#1\or
    \ensuremath{\dagger}\or
    \ensuremath{\ddagger}\or
    \ensuremath{\mathsection}\or
    \ensuremath{\mathparagraph}\or
    \ensuremath{\|}\or
    \ensuremath{\dagger\dagger}\or
    \ensuremath{\ddagger\ddagger}\or
    \ensuremath{\mathsection\mathsection}\or
    \ensuremath{\mathparagraph\mathparagraph}\else
    \@ctrerr
  \fi}
\makeatother

\begin{document}
\maketitle

\begin{abstract}
We study finite-rank multiplicative deformations of rotationally invariant non-Hermitian random matrices. More precisely, we consider models of the form $\vb{A}(\vb{I}+\vb{T})$, where $\vb{A}$ is a large rotationally invariant non-Hermitian random matrix, $\vb{T}$ is a finite-rank normal perturbation, and $\vb{I}$ denotes the identity matrix. We characterize the emergence of outlier eigenvalues, their fluctuations, and the associated eigenvector overlaps. Our results provide a multiplicative non-Hermitian counterpart to the classical Baik--Ben Arous--Péché framework.
\end{abstract}

In high-dimensional random matrix theory, multiplicative finite-rank deformations provide a natural framework for studying how a low-dimensional signal may affect the spectrum of a large random operator. A typical model is $\vb{A}(\vb{I}+\vb{T})$, where $\vb{T}$ is normal and has fixed rank as $N\to\infty$. If $\vb{A}$ is Hermitian positive, one may equivalently work with the symmetrized deformation
\begin{equation}
\vb{A}^{1/2}(\vb{I}+\vb{T})\vb{A}^{1/2},
\end{equation}
which has the same nonzero eigenvalues as $\vb{A}(\vb{I}+\vb{T})$. In this setting, the main question is whether the finite-rank deformation can create eigenvalues outside the limiting spectral support of the unperturbed matrix, and, when this happens, how the corresponding eigenvectors and fluctuations behave.

In the Hermitian setting, this question is closely related to spiked covariance models and to the Baik--Ben Arous--Péché transition. In such models, a finite-rank perturbation may create eigenvalues which detach from the Marchenko--Pastur bulk. Below the critical threshold, no outlier appears and the eigenvectors remain asymptotically orthogonal to the spike directions. Above the threshold, deterministic outliers emerge, with Gaussian fluctuations and non-trivial eigenvector overlaps~\cite{BBP2005,BaikSilverstein2006,Paul2007,BaiYao2008,BenaychNadakuditi2011,DingJi2023}.

More generally, multiplicative finite-rank deformations of Hermitian or positive random matrices can be described through free probability and subordination functions. In invariant models, the limiting location of possible outliers is governed by the multiplicative analogue of the resolvent equation appearing in the additive theory. The theory of outliers and eigenvectors for invariant multiplicative models is now well established in the Hermitian setting~\cite{BelinschiBercoviciCapitaineFevrier2017,DingJi2023, FirstCourseRMT}.

The non-Hermitian multiplicative problem is more delicate. Indeed,
\begin{equation}
    \vb{A}(\vb{I}+\vb{T})
    =
    \vb{A}+\vb{A}\vb{T},
\end{equation}
so the perturbation $\vb{A}\vb{T}$ is random and strongly correlated with $\vb{A}$, rather than being a deterministic finite-rank deformation. Consequently, the criteria developed for additive low-rank perturbations do not apply directly. To the best of our knowledge, the main available result in the non-Hermitian setting is due to Benaych-Georges and Rochet~\cite{BenaychRochet2017}, who showed that, in the bi-invariant setting\footnote{We say that a random matrix~$\vb{M}$ is bi-invariant if $\vb{M}$ and $\vb{U}\vb{M}\vb{V}$ have the same distribution for all unitary matrices~$\vb{U}$ and~$\vb{V}$.}, such multiplicative finite-rank deformations do not create outliers outside the limiting ring. Related questions also appear in the case where the unperturbed matrix is unitary ~\cite{Fyodorov2001,FyodorovSommers2003, DubachReker2024}.

The aim of this paper is to develop a framework for studying multiplicative finite-rank deformations of rotationally invariant non-Hermitian random matrices\footnote{In the sense that $\vb{A}$ and $\vb{U}\vb{A}\vb{U}^*$ have the same distribution for every unitary matrix $\vb{U}$.}. We investigate when a deformation of the form $\vb{A}(\vb{I}+\vb{T})$ can create eigenvalues outside the limiting spectral support of the unperturbed matrix. We also study the associated eigenvectors, as well as the fluctuations of the outliers when they are well separated from the bulk. The results presented here extend previous analyses and rely extensively on the theory developed in~\cite{bousseyroux1}.

\section{General setting and main result}
Let $\vb{A}$ be a large, rotationally invariant random matrix, not necessarily Hermitian, such that $\vb{U}\vb{A}\vb{U}^*$ has the same distribution as $\vb{A}$ for any unitary matrix $\vb{U}$. Let $\vb{T}$ be a finite-rank normal perturbation, meaning that there exist an integer $n$ and complex scalars $s_1, \ldots, s_n \in \C$, as well as unit vectors $u_1, \ldots, u_n \in \C^N$, such that
\begin{equation}
    \vb{T} = \sum_{j=1}^n s_j\,u_j\,u_j^*,
\end{equation}
where $u^* := \overline{u}^{\,T}$ for any $u \in \C^N$. The main analytical tool will be the Stieltjes transform $\g_{\vb{A}}$ of $\vb{A}$, defined in the large-$N$ limit as
\begin{equation}
    \g_{\vb{A}}(z)
    \;:=\;
    \lim_{N\to+\infty} \frac{1}{N}\Tr\!\bigl[(z\vb{I} - \vb{A})^{-1}\bigr],
    \qquad z \in \C.
\end{equation}
More generally, we write
\begin{equation}
    \tau(\vb{B})
    :=
    \lim_{N\to+\infty}\frac{1}{N}\Tr\vb{B}
\end{equation}
whenever this limit exists. We also use the $\mathfrak{t}$-transform, defined by
\begin{equation}
    \mathfrak{t}_{\vb{A}}(z)
    =
    \tau\!\left(\vb{A}(z\vb{I} - \vb{A})^{-1}\right)
    =
    z\g_{\vb{A}}(z) - 1.
\end{equation}

We will also make use of the $\mathcal{R}_{1,\vb{A}}$ and $\mathcal{R}_{2,\vb{A}}$ transforms associated with $\vb{A}$, as well as of their corresponding multivalued functions $\widetilde{\mathcal{R}}_{1,\vb{A}}$ and $\widetilde{\mathcal{R}}_{2,\vb{A}}$, introduced in~\cite{bousseyroux1}. We refer the reader to that reference, since the proofs presented here rely extensively on it. We denote by $\rho_{\vb{A}}$ the limiting spectral distribution of $\vb{A}$ and by $\supp(\rho_{\vb{A}})$ its support. An outlier of $\vb{A}(\vb{I}+\vb{T})$ denotes an eigenvalue lying in $\C\setminus\supp(\rho_{\vb{A}})$. We denote by $\Omega_{\vb{A}}$ the set of points $z\in \C\setminus\supp(\rho_{\vb{A}})$ such that, in a neighbourhood of $z$, $\g_{\vb{A}}$ is identically zero.

For $z\in \C\setminus\big(\supp(\rho_{\vb{A}})\cup\Omega_{\vb{A}}\big)$, we recall Result~4 of~\cite{bousseyroux1}, which states that
\begin{equation}\label{eq1}
    \partial_\alpha \mathcal{R}_{1,\vb{A}}(0,\g_{\vb{A}}(z))
    =
    \frac{1}{|\g_{\vb{A}}(z)|^2}
    -
    \frac{1}{h_{\vb{A}}(z)},
\end{equation}
and
\begin{equation}\label{eq2}
    \g_{\vb{A}}(z)
    =
    \frac{1}{z - \mathcal{R}_{2,\vb{A}}(0,\g_{\vb{A}}(z))},
\end{equation}
where
\begin{equation}\label{defh}
    h_{\vb{A}}(z)
    =
    \lim_{N\to +\infty}
    \frac{1}{N}\Tr\!\left([(z\vb{I} - \vb{A})(z\vb{I} - \vb{A})^*]^{-1}\right).
\end{equation}
The functions $\mathcal{R}_{1,\vb{A}}$ and $\mathcal{R}_{2,\vb{A}}$ denote appropriate branches of
$\widetilde{\mathcal{R}}_{1,\vb{A}}$ and $\widetilde{\mathcal{R}}_{2,\vb{A}}$, respectively.

Furthermore, in~\cite{bousseyroux4}, we extended the classical $S$-transform to the non-Hermitian setting. This transform satisfies the analogue of Eq.~\eqref{eq2}:
\begin{equation}\label{relationS}
    S_{\vb{A}}(\mathfrak{t}_{\vb{A}}(z))
    =
    \frac{\mathfrak{t}_{\vb{A}}(z) + 1}{\mathfrak{t}_{\vb{A}}(z) z},
\end{equation}
where $S_{\vb{A}}$ denotes a suitable non-Hermitian $S$-transform of $\vb{A}$. A definition that will be useful in the sequel was given in~\cite{bousseyroux4}:
\begin{equation}\label{defk}
    k^*(\vb{A})
    =
    \min \left\{
    k \in \N^*
    \,:\,
    \tau\!\left(\vb{A}^k\right) \neq 0
    \right\},
\end{equation}
with values in $\N^* \cup \{+\infty\}$. For a Hermitian matrix $\vb{H}$, one has $k^*(\vb{H}) = +\infty$ if $\vb{H}$ is the zero matrix, $k^*(\vb{H}) = 2$ if $\vb{H}$ is centered and non-zero, and $k^*(\vb{H}) = 1$ otherwise.

We can now state the two main results. They naturally split according to whether $s_j=-1$ or not. The proof is given in the appendices in the rank-one case $n=1$.

\begin{result}\label{main_result}

Fix $1 \le j \le n$ such that $s_j\neq -1$. In the high-dimensional limit, the rank-one perturbation $s_j\,u_j\,u_j^*$ may produce potential outliers of $\vb{M}:=\vb{A} (\vb{I}+ \vb{T})$, which are solutions of
\begin{equation}\label{solution_outlier}
    \mathfrak{t}_{\vb{A}}(z) = \frac{1}{s_j}.
\end{equation}

\begin{enumerate}
\item[\textbf{(i)}] \textbf{Outlier location.}
One can equivalently write
\begin{equation}\label{formula_outlier}
z_j
=
\frac{s_j+1}{S_{\vb{A}}(1/s_j)},
\end{equation}
where $S_{\vb{A}}$ is an $S$-transform of $\vb{A}$.

\item[\textbf{(ii)}] \textbf{Eigenvector alignment.}  
Let $\phi_j$ be a normalized eigenvector associated with the outlier $z_j$. Then, in the high-dimensional limit,

\begin{equation}
    \bigl|\langle u_j,\phi_j\rangle\bigr|^2
    \underset{N\to+\infty}{\longrightarrow}
    \frac{
    \left|
    \tau\!\left((z_j\vb{I}-\vb{A})^{-1}\vb{A}\right)
    \right|^2
    }
    {
    \tau\!\left(
    \vb{A}^*
    \left[(z_j\vb{I}-\vb{A})(z_j\vb{I}-\vb{A})^*\right]^{-1}
    \vb{A}
    \right)
    },
\end{equation}
which can be written as

\begin{equation}\label{overlap}
    \bigl|\langle u_j,\phi_j\rangle\bigr|^2
    \underset{N\to+\infty}{\longrightarrow}
    1
    -
    \left|
        \frac{1+1/s_j}{S_{\vb{A}}(1/s_j)}
    \right|^2
    \,
    \partial_{\alpha}\mathcal{R}_{1,\vb{A}^{-1}}
    \left(
        0,
        -\frac{1+1/s_j}{S_{\vb{A}}(1/s_j)}
    \right).
\end{equation}
Here, $\mathcal{R}_{1,\vb{A}^{-1}}$ denotes a determination of $\widetilde{\mathcal{R}}_{1,\vb{A}^{-1}}$.

\item[\textbf{(iii)}] \textbf{Fluctuations.}  
For large $N$, the fluctuations of the outlier position $z_j$ are Gaussian with variance $\sigma_j^2/N$, where

\begin{equation}
    \sigma_j^2 = \frac{
        \tau\!\left(
            \vb{A}^*
    \left[(z_j\vb{I}-\vb{A})(z_j\vb{I}-\vb{A})^*\right]^{-1}
    \vb{A}
        \right)
        -
        \left|
            \tau\!\left(
                (z_j \vb{I} - \vb{A})^{-1}\vb{A}
            \right)
        \right|^2
    }{
        \left|
            \tau(\vb{A}(z_j\vb{I} - \vb{A})^{-2})
        \right|^2
    },
\end{equation}
which can be written as

\begin{equation}\label{express_variance}
    \sigma_j^2
    =
    \frac{
        \partial_{\alpha}
    \mathcal{R}_{1,\vb{A}^{-1}}
    \!\left(
        0,
        -\frac{1+1/s_j}{S_{\vb{A}}(1/s_j)}
    \right)
    }{
        |s_j|^2
        \left(
            1
            -
            \partial_{\alpha}
            \mathcal{R}_{1,\vb{A}^{-1}}
            \!\left(
                0,
                -\frac{1+1/s_j}{S_{\vb{A}}(1/s_j)}
            \right)
            \left|
                \frac{1+1/s_j}{S_{\vb{A}}(1/s_j)}
            \right|^2
        \right)
    }
    \left|
        \frac{
            s_j^2 S_{\vb{A}}(1/s_j)
            +
            (s_j+1)S_{\vb{A}}'(1/s_j)
        }{
            S_{\vb{A}}(1/s_j)^2
        }
    \right|^2.
\end{equation}
where $\mathcal{R}_{1,\vb{A}^{-1}}$ denotes a determination of
$\widetilde{\mathcal{R}}_{1,\vb{A}^{-1}}$ and $S_{\vb{A}}$ is an
$S$-transform of $\vb{A}$.

\end{enumerate}

\end{result}

\begin{remarks}

    \item Let $z\in \Omega_{\vb{A}}$. Then $\mathfrak{t}_{\vb{A}}(z)=-1$. Therefore, if $s_j\neq -1$, the outliers described by the theorem cannot lie in $\Omega_{\vb{A}}$.
    \item If $\vb{A}$ is simply the matrix $\alpha \vb{I}_N$, then
    $S_{\vb{A}}(g)=1/\alpha$, and we recover that the outliers are given by
    \begin{equation}
        z_j=\alpha(s_j+1).
    \end{equation}

    \item The number of outliers is bounded by the number of $S$-transforms of $\vb{A}$.

    \item The expressions simplify remarkably when $\vb{A}$ is Hermitian. Indeed, as recalled in~\cite{bousseyroux1}, in this case one has
    \begin{equation}
        \partial_{\alpha} \mathcal{R}_{1,\vb{A}}(0,g)
        =
        \frac{\Im(R_{\vb{A}}(g))}{\Im(g)},
    \end{equation}
    where $R_{\vb{A}}$ is the classical $R$-transform in the Hermitian case. We also know from~\cite{FirstCourseRMT} that
    \begin{equation}
        S_{\vb{A}^{-1}}(x)
        =
        \frac{1}{S_{\vb{A}}(-x-1)}.
    \end{equation}

    Using these two relations, we can write
    \begin{equation}
    \left|
        \frac{1+1/s_j}{S_{\vb{A}}(1/s_j)}
    \right|^2
    \,
    \partial_{\alpha}\mathcal{R}_{1,\vb{A}^{-1}}
    \left(
        0,
        -\frac{1+1/s_j}{S_{\vb{A}}(1/s_j)}
    \right)
    =
    \left|uS_{\vb{A}^{-1}}(u)\right|^2
    \frac{
        \Im\!\left(R_{\vb{A}^{-1}}\!\left(uS_{\vb{A}^{-1}}(u)\right)\right)
    }{
        \Im\!\left(uS_{\vb{A}^{-1}}(u)\right)
    },
    \end{equation}
    where
    \begin{equation}
        u=-1-\frac{1}{s_j}.
    \end{equation}

    We then use the definition of the $S$-transform in~\cite{bousseyroux4}, which gives
    \begin{equation}
        R_{\vb{A}^{-1}}\!\left(uS_{\vb{A}^{-1}}(u)\right)
        =
        \frac{1}{S_{\vb{A}^{-1}}(u)}.
    \end{equation}

    Putting everything together, we obtain
    \begin{equation}
    \left|
        \frac{1+1/s_j}{S_{\vb{A}}(1/s_j)}
    \right|^2
    \,
    \partial_{\alpha}\mathcal{R}_{1,\vb{A}^{-1}}
    \left(
        0,
        -\frac{1+1/s_j}{S_{\vb{A}}(1/s_j)}
    \right)
    =
    \frac{
        \Im\!\left(S_{\vb{A}}(1/s_j)\right)
    }{
        \Im\!\left(
            \frac{s_j S_{\vb{A}}(1/s_j)}{1+s_j}
        \right)
    }.
    \end{equation}

    Hence the overlap is given by
    \begin{equation}\label{eqH}
        \frac{
            \Im\!\left(
                \frac{S_{\vb{A}}(1/s_j)}{1+s_j}
            \right)
        }{
            \Im\!\left(
                \frac{s_j S_{\vb{A}}(1/s_j)}{1+s_j}
            \right)
        }.
    \end{equation}

    \item Using formula~\eqref{overlap}, we conjecture that the transition occurs when a point $s\in\C$ crosses the curve defined by
    \begin{equation}
        \left|
            \frac{1+1/s}{S_{\vb{A}}(1/s)}
        \right|^2
        \,
        \partial_{\alpha}\mathcal{R}_{1,\vb{A}^{-1}}
        \left(
            0,
            -\frac{1+1/s}{S_{\vb{A}}(1/s)}
        \right)
        =
        1.
    \end{equation}
    Here $\mathcal{R}_{1,\vb{A}^{-1}}$ denotes a determination of
    $\widetilde{\mathcal{R}}_{1,\vb{A}^{-1}}$. This provides a natural extension of the BBP transition threshold.

    In the Hermitian case, using formula~\eqref{eqH}, this condition reduces to
    \begin{equation}\label{thre}
        \frac{S_{\vb{A}}(1/s)}{1+s}\in\R.
    \end{equation}
    Thus, in this case, an outlier becomes visible when $s\in\C$ crosses the curve defined by Eq.~\eqref{thre}.

    \item Assume first that $k=k^*(\vb{A})<+\infty$. When $z$ tends to infinity, one has
\begin{equation}
\mathfrak{t}_{\vb{A}}(z)
\underset{z\to\infty}{\sim}
\frac{\tau(\vb{A}^{k})}{z^{k}}.
\end{equation}
Thus, when $|s_j|\to+\infty$, the equation~\eqref{solution_outlier} has
asymptotically $k$ solutions, given by the $k$-th roots of
\begin{equation}
\tau(\vb{A}^{k})s_j.
\end{equation}
Hence, there are $k$ outliers, whose typical order of magnitude is
$|s_j|^{1/k}$. Moreover, for large $|s_j|$, the overlaps are asymptotically given by
\begin{equation}\label{ofinifni}
\frac{|\tau(\vb{A})|^2}{\tau(\vb{A}\vb{A}^*)}.
\end{equation}

Furthermore, the corresponding variances are given by
\begin{equation}\label{fluctuinfini}
\frac{|s_j|^2 (\tau(\vb{A}\vb{A}^*) - |\tau(\vb{A})|^2)}{k^2}.
\end{equation}

Let us note that, contrary to the additive case studied
in~\cite{bousseyroux2}, the variance in~\eqref{fluctuinfini} diverges
as $|s_j|\to\infty$. 

    \item As recalled for the Wigner case in~\cite{rao2007multiplication} and also in~\cite{bousseyroux4}, there are two possible $S$-transforms for the Wigner matrix, namely
    \begin{equation}
        S_{\vb{W}}(t)=\frac{\pm 1}{\sqrt{t}}.
    \end{equation}
    We therefore obtain two possible outliers, given by
    \begin{equation}\label{wigner_case}
        z=\pm\frac{s+1}{\sqrt{s}}.
    \end{equation}

    These outliers become visible when $|s|\geq 1$. Indeed, applying Eq.~\eqref{thre} gives the condition
    \begin{equation}
        \frac{\sqrt{s}}{1+s}\in\R.
    \end{equation}
    Equivalently, this transition curve is given by $|s|=1$. Moreover, the corresponding overlaps tend to $0$ as $|s|\to+\infty$, and the Gaussian fluctuations have asymptotic variance
    \begin{equation}
        \frac{|s|^2}{4},
    \end{equation}
    as follows from Eqs.~\eqref{ofinifni} and~\eqref{fluctuinfini}.

    \item Figure~\ref{fig:triangle} shows a case where $k^*(\vb{A})=3$.
    To obtain this example, we use a uniform distribution on a triangle.
    We then observe three possible outliers, with large fluctuations.

    \begin{figure}
        \centering
        \includegraphics[width=0.5\linewidth]{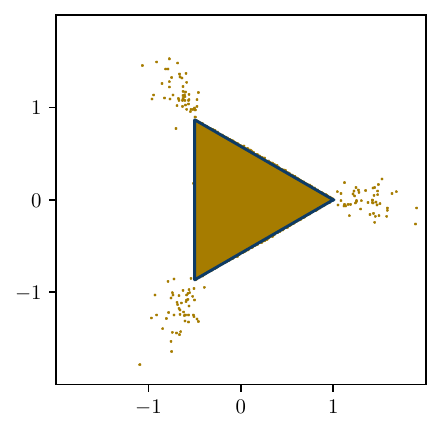}
        \caption{Empirical eigenvalues of $\vb{A}(\vb{I}_N+suu^*)$, with $N=200$ and $s=20$, where $u$ is a unit vector. We set
$\vb{A}=\vb{U}\vb{D}\vb{U}^*$, where $\vb{U}$ is Haar-distributed and independent of the diagonal matrix
$\vb{D}=\operatorname{diag}(Z_1,\ldots,Z_N)$, whose entries are sampled uniformly from the equilateral triangle with vertices $1$, $e^{2i\pi/3}$, and $e^{4i\pi/3}$. Eigenvalues from $100$ independent realizations are superimposed. The blue contour shows the boundary of the limiting triangular support.
}
        \label{fig:triangle}
    \end{figure}

    \item One may expect that, when $|s|$ goes to infinity,
    \begin{equation}\label{approximation}
        \vb{A}(\vb{I}+s u u^*)
        \approx
        s\,\vb{A}u u^*.
    \end{equation}
    This would suggest that the outlier is given by
    $s\langle u,\vb{A}u\rangle$, with eigenvector proportional to
    $\vb{A}u$. This is indeed what we recover in the case
    $k^*(\vb{A})=1$. However, when $\tau(\vb{A})=0$, so that
    $k^*(\vb{A})\geq 2$, the approximation~\eqref{approximation} no longer
    captures the limiting outliers. This provides an example of
    non-commutation between the limits $N\to+\infty$ and
    $|s|\to+\infty$: if one first takes $|s|\to+\infty$ at fixed $N$,
    the approximation~\eqref{approximation} is valid, whereas the
    high-dimensional limit leads to a different behaviour.

    \item Let us now consider the case where $\vb{A}$ is a large
    bi-invariant matrix. Using the single ring theorem~\cite{feinberg1997non,haagerup2000brown,guionnet2011single},
    we know that the support of the limiting spectral distribution of
    $\vb{A}$ is a ring, with inner and outer radii denoted by
    $r_{-,\vb{A}}$ and $r_{+,\vb{A}}$, respectively. Furthermore, by
    elementary electrostatics and the rotational invariance of the
    distribution in the complex plane, one can show that
    \begin{equation}\label{gbi}
        \g_{\vb{A}}(z)
        =
        \begin{cases}
            0, & \text{if } |z|<r_{-,\vb{A}}, \\
            \dfrac{1}{z}, & \text{if } |z|>r_{+,\vb{A}}.
        \end{cases}
    \end{equation}
    Therefore,
    \begin{equation}
        \mathfrak{t}_{\vb{A}}(z)
        =
        \begin{cases}
            -1, & \text{if } |z|<r_{-,\vb{A}}, \\
            0, & \text{if } |z|>r_{+,\vb{A}}.
        \end{cases}
    \end{equation}
    We deduce that, for $s_j\neq -1$, no outlier is created outside the
    limiting ring~\cite{BenaychRochet2017}. This is illustrated in Fig.~\ref{fig:bruit}.
\end{remarks}
\begin{result}\label{res2}
Fix $1\leq j\leq n$ and assume that $s_j=-1$. Then the following statements hold.

\begin{enumerate}[label=\textnormal{(\roman*)}]
    \item \textbf{Deterministic eigenvalue at zero.}
    The perturbation produces a deterministic eigenvalue at $0$, with associated right eigenvector $u_j$.

    Any additional nonzero outlier is, at finite $N$, a zero of
    \begin{equation}\label{eq:finite_N_s_minus_one}
        z\longmapsto
        \langle u_j,(z\vb{I}-\vb{A})^{-1}u_j\rangle.
    \end{equation}

    \item \textbf{Isolated outliers outside $\Omega_{\vb{A}}$.}
    Let
    \begin{equation}
        z_*
        \in
        \C\setminus
        \bigl(\supp(\rho_{\vb{A}})\cup\Omega_{\vb{A}}\bigr)
    \end{equation}
    be the limiting location of a nonzero outlier. Then necessarily
    \begin{equation}\label{eq:zero_g_outlier}
        \g_{\vb{A}}(z_*)=0.
    \end{equation}

    If $\phi$ denotes the associated normalized right eigenvector, then
    \begin{equation}\label{eq:overlap_s_minus_one}
        |\langle u_j,\phi\rangle|^2
        \underset{N\to+\infty}{\longrightarrow}
        \frac{1}
        {1+|z_*|^2 h_{\vb{A}}(z_*)},
    \end{equation}
    where $h_{\vb{A}}$ is defined in Eq.~\eqref{defh}.
    Equivalently,
    \begin{equation}
        |\langle u_j,\phi\rangle|^2
        \underset{N\to+\infty}{\longrightarrow}
        1-|z_*|^2\,
        \partial_{\alpha}
        \mathcal{R}_{1,\vb{A}^{-1}}(0,z_*).
    \end{equation}

    If, in addition, $z_*$ is a simple zero of $\g_{\vb{A}}$, then the outlier has Gaussian fluctuations of order $N^{-1/2}$. More precisely, its variance is $\sigma_*^2/N$, where
    \begin{equation}\label{eq:variance_s_minus_one}
        \sigma_*^2
        =
        \frac{h_{\vb{A}}(z_*)}
        {|\g_{\vb{A}}'(z_*)|^2}.
    \end{equation}

    \item \textbf{Outliers inside $\Omega_{\vb{A}}$.}
    Inside $\Omega_{\vb{A}}$, the function $\g_{\vb{A}}$ vanishes identically, so that Eq.~\eqref{eq:zero_g_outlier} no longer selects isolated deterministic locations. Instead, for large $N$, the nonzero outliers lying in $\Omega_{\vb{A}}$ are described by the limiting mean outlier intensity
    \begin{equation}\label{desrho}
        \varrho_{\vb{A}}^{\rm out}(z)
        =
        \frac{1}{4\pi}\Delta\log h_{\vb{A}}(z),
        \qquad
        z\in\Omega_{\vb{A}},
    \end{equation}
    where
    \begin{equation}
        \Delta
        =
        \partial_x^2+\partial_y^2
        =
        4\partial_z\partial_{\bar z}
    \end{equation}
    denotes the Laplacian.

    Moreover, if $\phi$ denotes the normalized right eigenvector associated with a nonzero outlier at $z\in\Omega_{\vb{A}}$, then
    \begin{equation}\label{overbi}
        |\langle u_j,\phi\rangle|^2
        \underset{N\to+\infty}{\longrightarrow}
        \frac{1}
        {1+|z|^2 h_{\vb{A}}(z)}
        =
        1-|z|^2\,
        \partial_{\alpha}
        \mathcal{R}_{1,\vb{A}^{-1}}(0,z).
    \end{equation}
\end{enumerate}
\end{result}
\begin{remarks}
    \item According to~\cite{bousseyroux3}, the overlap in Eq.~\eqref{overbi} vanishes at the boundary of the spectrum.

    \item Since the distribution of $\vb{A}$ is invariant under rotations, the distribution of the zeros of
    \begin{equation}
        z\mapsto \langle u_j,(z\vb{I}-\vb{A})^{-1}u_j\rangle
    \end{equation}
    is the same as the distribution of the zeros of
    \begin{equation}
        z\mapsto \big[(z\vb{I}-\vb{A})^{-1}\big]_{11}.
    \end{equation}
    By the Schur complement formula, these zeros are precisely the eigenvalues of the lower-right principal block of $\vb{A}$ of size $(N-1)\times(N-1)$. Thus, the outlier intensity $\varrho_{\vb{A}}^{\rm out}$ also describes, in the high-dimensional limit, the mean intensity of the eigenvalues of this principal block inside $\Omega_{\vb{A}}$.

    \item Let $\vb{A}$ be a bi-invariant matrix. As recalled in Eq.~\eqref{gbi}, one has
    \begin{equation}
        \Omega_{\vb{A}}
        =
        \{z\in\C:\ |z|<r_{-,\vb{A}}\}.
    \end{equation}
    Moreover, for $z\in\Omega_{\vb{A}}$,
    \begin{equation}
        h_{\vb{A}}(z)
        =
        \frac{1}{r_{-,\vb{A}}^2-|z|^2}.
    \end{equation}
    Therefore,
    \begin{equation}
        \varrho_{\vb{A}}^{\rm out}(z)
        =
        \frac{1}{4\pi}
        \Delta\log h_{\vb{A}}(z)
        =
        \frac{1}{4\pi}
        \Delta\left[-\log\left(r_{-,\vb{A}}^2-|z|^2\right)\right].
    \end{equation}
    Since
    \begin{equation}
        \Delta\left[-\log\left(r_{-,\vb{A}}^2-|z|^2\right)\right]
        =
        \frac{4r_{-,\vb{A}}^2}
        {\left(r_{-,\vb{A}}^2-|z|^2\right)^2},
    \end{equation}
    we obtain
    \begin{equation}
        \varrho_{\vb{A}}^{\rm out}(z)
        =
        \frac{r_{-,\vb{A}}^2}
        {\pi\left(r_{-,\vb{A}}^2-|z|^2\right)^2}.
    \end{equation}
    Denoting by $N_{\vb{A}}^{\rm out}(r)$ the number of nonzero outliers in the disk $\{|z|<r\}$, we obtain, for $r<r_{-,\vb{A}}$,
\begin{equation}\label{eq:count_biinvariant}
    \mathbb{E}\,N_{\vb{A}}^{\rm out}(r)
    =
    \int_{|z|<r}
    \varrho_{\vb{A}}^{\rm out}(z)\,d^2z
    =
    \frac{r^2}{r_{-,\vb{A}}^2-r^2}.
\end{equation}
In particular, in the large-$N$ limit, this quantity no longer depends on $N$.
    Moreover, Eq.~\eqref{overbi} gives, in this bi-invariant case,
    \begin{equation}\label{eq:overlap_biinvariant}
        |\langle u_j,\phi\rangle|^2
        =
        1-\left|\frac{z}{r_{-,\vb{A}}}\right|^2.
    \end{equation}

We test these formulas in Fig.~\ref{fig:bruit}. 

 \item Equation~\eqref{eq:count_biinvariant} coincides with the large-$N$ limit of the classical formula for the
$(N-1)\times(N-1)$ truncation of a Haar unitary matrix: denoting by $z_1,\ldots,z_{N-1}$ the eigenvalues of this principal truncation, one has, for $r<1$ fixed,
\begin{equation}
    \mathbb{E}\,\#\{k:\ |z_k|\leq r\}
    =
    \frac{r^2(1-r^{2N-2})}{1-r^2}
    \underset{N\to+\infty}{\longrightarrow}
    \frac{r^2}{1-r^2},
\end{equation}
see~\cite{zyczkowski2000truncations,petz2005large}.
\end{remarks}

    \begin{figure}
    \centering
    \includegraphics[width=0.9\linewidth]{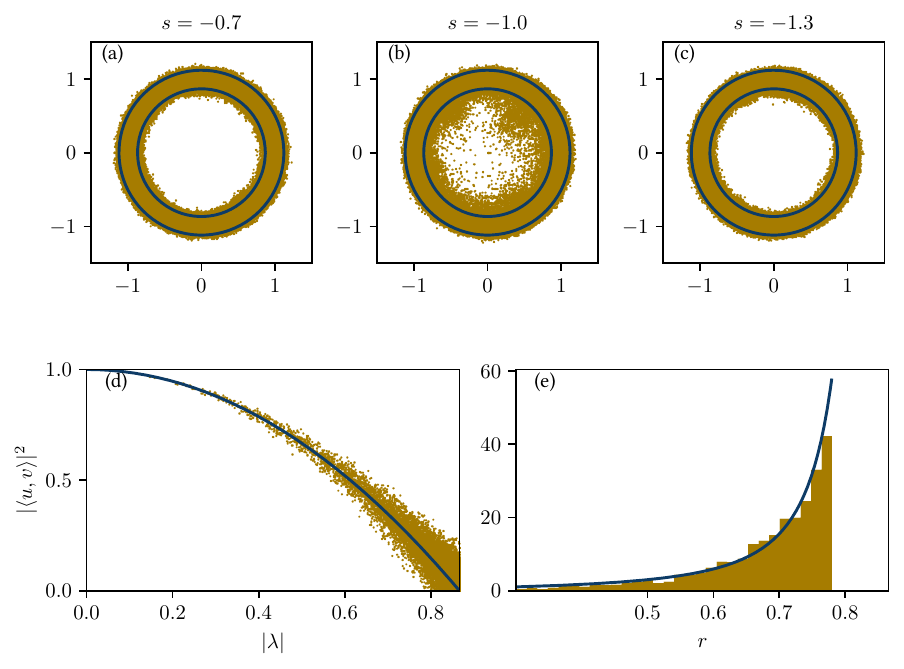}
    \caption{Spectrum of the rank-one multiplicative perturbation
$\vb{M}=(\vb{A}+\sigma \vb{G})(\vb{I}_N+suu^*)$,
where $\vb{A}$ is Haar unitary, $\vb{G}$ is a complex Ginibre matrix, $\sigma=0.5$, and $N=100$. Panels~(a)--(c) show the eigenvalues for $s=-0.7$, $-1$, and $-1.3$, respectively. The blue circles indicate the radii $\sqrt{1-\sigma^2}$ and $\sqrt{1+\sigma^2}$. Eigenvalues from $5000$ independent realizations are superimposed. Panel~(d) shows, for $s=-1$, the squared overlap $|\langle u,\phi\rangle|^2$ between a right eigenvector $\phi$ and the spike direction $u$. The blue curve is the theoretical prediction from Eq.~\eqref{eq:overlap_biinvariant}. Panel~(e) shows the radial profile of the nonzero outliers, obtained from the histogram of the moduli $|\lambda|$ restricted to $|\lambda|<\sqrt{1-\sigma^2}$, after removing the deterministic zero eigenvalue. The blue curve is the derivative with respect to $r$ of Eq.~\eqref{eq:count_biinvariant}.}
    \label{fig:bruit}
\end{figure}

\paragraph*{Acknowledgements.} 

This research was conducted within the Econophysics \& Complex Systems Research Chair, under the aegis of the Fondation du Risque, the Fondation de l’Ecole polytechnique, the Ecole polytechnique, and Capital Fund Management.

\appendix

\newpage

\section{Proofs}
We consider a large rotationally invariant random matrix~$\vb{A}$, and set $\vb{M} = \vb{A}(\vb{I} + s\,u u^*)$, where $s \in \C$ and $u \in \C^N$ is a unit vector. For any matrix~$\vb{B}$, we denote by $\vb{G}_{\vb{B}}(z)$ the resolvent of~$\vb{B}$ evaluated at a point $z \in \C$, defined by
\begin{equation}
    \vb{G}_{\vb{B}}(z) = (z\vb{I} - \vb{B})^{-1},
\end{equation}
where $\vb{I}$ denotes the identity matrix. We recall some relations that will be useful, proved in~\cite{bousseyroux1} and~\cite{bousseyroux4}. They are given by
\begin{equation}
    S_{\vb{A}^{-1}}(t) = \frac{1}{S_{\vb{A}}(-t-1)}
\end{equation}
and
\begin{equation}
    \partial_{\alpha} \mathcal{R}_{1, \alpha\vb{A}}(0, g)
    =
    |\alpha|^2 \partial_{\alpha} \mathcal{R}_{1, \vb{A}}(0, \alpha g).
\end{equation}
Moreover,
\begin{equation}
    \mathcal{R}_{2, \alpha \vb{A}}(0, g)
    =
    \alpha \mathcal{R}_{2, \vb{A}}(0, \alpha g).
\end{equation}

Another ingredient in the proof of the main result will be to notice that $z$ is an outlier of~$\vb{M}$ if and only if $1$ is an outlier of
\begin{equation}\label{equivalence}
    z\vb{A}^{-1} - s u u^*,
\end{equation}
and the associated eigenvector is the same. We are therefore reduced to an additive BBP problem, and we may apply the results of~\cite{bousseyroux2}.

\subsection{Proof of (i)}\label{sec:outlier}

We apply the first point of the main result of~\cite{bousseyroux2}, which states that $1$ is an outlier of $z\vb{A}^{-1} - s uu^*$ if and only if
\begin{equation}
    \g_{z\vb{A}^{-1}}(1) = \frac{-1}{s},
\end{equation}
that is,
\begin{equation}\label{t_value}
    \mathfrak{t}_{\vb{A}}(z) = \frac{1}{s}.
\end{equation}
Then, to obtain~\eqref{formula_outlier}, it is enough to use~\eqref{relationS} and~\eqref{t_value}. We thus obtain the first statement of the result.

\subsection{Proof of (ii)}\label{sec:eigenvectors}

Let $z$ be a possible outlier of $\vb{M}=\vb{A}(\vb{I}+s\,u\,u^*)$ and let $\phi$ be a unit right eigenvector associated with it.

We have
\begin{equation}
    \vb{A}(\vb{I}+s\,u\,u^*)\phi
    =
    z\phi,
\end{equation}
that is,
\begin{equation}
    \vb{A}\phi
    +
    s\,\langle u,\phi\rangle\,\vb{A}u
    =
    z\phi.
\end{equation}
Thus
\begin{equation}
    (z\vb{I}-\vb{A})\phi
    =
    s\,\langle u,\phi\rangle\,\vb{A}u.
\end{equation}
It follows that $\phi$ is proportional to
\begin{equation}
    (z\vb{I}-\vb{A})^{-1}\vb{A}u.
\end{equation}
Equivalently, using
\begin{equation}
    (z\vb{I}-\vb{A})^{-1}\vb{A}
    =
    z(z\vb{I}-\vb{A})^{-1}-\vb{I},
\end{equation}
one can write this vector as
\begin{equation}
    \left(z(z\vb{I}-\vb{A})^{-1}-\vb{I}\right)u.
\end{equation}

Since $\phi$ is a unit vector, we have
\begin{equation}
    \phi
    =
    \frac{(z\vb{I}-\vb{A})^{-1}\vb{A}u}
    {
    \sqrt{
    \left\langle
    u
    \middle|
    \vb{A}^*
    \left[(z\vb{I}-\vb{A})(z\vb{I}-\vb{A})^*\right]^{-1}
    \vb{A}
    \middle|
    u
    \right\rangle
    }
    }.
\end{equation}
Hence,
\begin{equation}
    |\langle u,\phi\rangle|^2
    =
    \frac{
    \left|
    \left\langle
    u
    \middle|
    (z\vb{I}-\vb{A})^{-1}\vb{A}
    \middle|
    u
    \right\rangle
    \right|^2
    }
    {
    \left\langle
    u
    \middle|
    \vb{A}^*
    \left[(z\vb{I}-\vb{A})(z\vb{I}-\vb{A})^*\right]^{-1}
    \vb{A}
    \middle|
    u
    \right\rangle
    }.
\end{equation}

Using the rotational invariance of $\vb{A}$, this becomes, in the large-$N$ limit,
\begin{equation}
    |\langle u,\phi\rangle|^2
    =
    \frac{
    \left|
    \tau\!\left((z\vb{I}-\vb{A})^{-1}\vb{A}\right)
    \right|^2
    }
    {
    \tau\!\left(
    \vb{A}^*
    \left[(z\vb{I}-\vb{A})(z\vb{I}-\vb{A})^*\right]^{-1}
    \vb{A}
    \right)
    }.
\end{equation}

If one wants to express this formula only in terms of transforms of $\vb{A}$ or $\vb{A}^{-1}$, one can use the following argument. The second point of the main result of~\cite{bousseyroux2} gives that the overlap between $u$ and $\phi$ is given by
\begin{equation}
    1 - \frac{\partial_{\alpha} \mathcal{R}_{1, z \vb{A}^{-1}}(0, -1/s)}{|s|^2}.
\end{equation}
This can be rewritten as
\begin{equation}\label{po}
    1 - \frac{|z|^2 \partial_{\alpha} \mathcal{R}_{1, \vb{A}^{-1}}(0, -z/s)}{|s|^2}.
\end{equation}
Moreover,
\begin{equation}
    \frac{z}{s}
    =
    \frac{1 + 1/s}{S_{\vb{A}}(1/s)}.
\end{equation}
We thus obtain the second point of the result.

\subsection{Proof of (iii)}\label{sec:fluctuations}
The Sherman--Morrison formula~\cite{sherman1950adjustment} yields
\begin{equation}\label{eq:resolvent_rank_one}
    \vb{G}_{\vb{M}}(z)
    =
    \vb{G}_{\vb{A}}(z)
    +
    s\,\frac{
        \vb{G}_{\vb{A}}(z)\vb{A}u u^*\vb{G}_{\vb{A}}(z)
    }{
        1 - s\,u^*\vb{G}_{\vb{A}}(z)\vb{A}u
    }.
\end{equation}
Thus, at finite $N$, the possible outliers are given by the solutions of
\begin{equation}\label{eq:finite_N_outlier_condition}
    \bra{u}\vb{G}_{\vb{A}}(z)\vb{A}\ket{u}
    =
    \frac{1}{s}.
\end{equation}

First, let $\vb{B}$ be a large rotationally invariant non-Hermitian matrix. Using Eq.~(32) from~\cite{bousseyroux1}, one obtains in particular that the fluctuations of a diagonal entry of $\vb{B}$ are Gaussian with variance
\begin{equation}\label{eq:diagonal_entry_variance}
    \frac{
        \tau(\vb{B}\vb{B}^*) - |\tau(\vb{B})|^2
    }{N},
\end{equation}
as $N \to +\infty$, where $\tau$ denotes the normalized trace.

The idea is to apply this result to a suitably chosen matrix. Returning to~\eqref{eq:resolvent_rank_one}, we have seen that, for finite $N$, the possible outliers are given by the solutions of
\begin{equation}\label{eq:outlier_condition_finite_N}
    \bra{u}\vb{G}_{\vb{A}}(z)\vb{A}\ket{u}
    =
    \frac{1}{s}.
\end{equation}
Moreover, as $N \to +\infty$, one has $z \to z_*$, where
\begin{equation}\label{eq:outlier_condition_limit}
    \mathfrak{t}_{\vb{A}}(z_*)
    =
    \frac{1}{s}.
\end{equation}
Expanding $z$ around $z_*$, one finds that the fluctuations of $z$ are governed by those of
\begin{equation}
    \bra{u}\vb{G}_{\vb{A}}(z_*)\vb{A}\ket{u}
\end{equation}
in the large-$N$ limit. In particular, using~\eqref{eq:diagonal_entry_variance} with
\begin{equation}
    \vb{B}
    =
    \vb{G}_{\vb{A}}(z_*)\vb{A},
\end{equation}
we obtain Gaussian fluctuations with variance $\sigma^2/N$, where
\begin{equation}\label{eq:variance_first_form}
    \sigma^2
    :=
    \frac{
        \tau\!\left(
            \vb{G}_{\vb{A}}(z_*)\vb{A}\vb{A}^*
            \vb{G}_{\vb{A}}(z_*)^*
        \right)
        -
        \left|
            \tau\!\left(
                \vb{G}_{\vb{A}}(z_*)\vb{A}
            \right)
        \right|^2
    }{
        \left|
            \partial_z \mathfrak{t}_{\vb{A}}(z_*)
        \right|^2
    }.
\end{equation}
Equivalently, this coefficient can be rewritten as
\begin{equation}\label{eq:variance_second_form}
    \sigma^2
    =
    \frac{
        h_{\vb{A}^{-1}}(1/z_*)
        -
        \left|
            \g_{\vb{A}^{-1}}(1/z_*)
        \right|^2
    }{
        |z_*|^2
        \left|
            \partial_z \mathfrak{t}_{\vb{A}}(z_*)
        \right|^2
    }.
\end{equation}

Using~\eqref{eq1}, \eqref{solution_outlier}, and the formula~\eqref{formula_outlier}, we obtain
\begin{equation}\label{eq:final_variance_intermediate}
    \sigma^2
    =
    \frac{
        1
    }{
        |s|^2
        \left(
            1
            -
            \partial_{\alpha}
            \mathcal{R}_{1,\vb{A}^{-1}}
            \!\left(
                0,
                -\frac{z_*}{s}
            \right)
            \left|
                \frac{z_*}{s}
            \right|^2
        \right)
    }
    \frac{
        \partial_{\alpha}
        \mathcal{R}_{1,\vb{A}^{-1}}
        \!\left(
            0,
            -\frac{z_*}{s}
        \right)
    }{
        \left|
            \partial_z \mathfrak{t}_{\vb{A}}(z_*)
        \right|^2
    }.
\end{equation}
Furthermore, using~\eqref{relationS}, one obtains
\begin{equation}\label{eq:derivative_t_relation}
    -\partial_z\mathfrak{t}_{\vb{A}}(z)
    =
    \frac{
        \mathfrak{t}_{\vb{A}}(z)^2
        S_{\vb{A}}\!\left(
            \mathfrak{t}_{\vb{A}}(z)
        \right)^2
    }{
        S_{\vb{A}}\!\left(
            \mathfrak{t}_{\vb{A}}(z)
        \right)
        +
        \mathfrak{t}_{\vb{A}}(z)
        \left(
            \mathfrak{t}_{\vb{A}}(z)+1
        \right)
        S_{\vb{A}}'\!\left(
            \mathfrak{t}_{\vb{A}}(z)
        \right)
    }.
\end{equation}
Returning to~\eqref{eq:final_variance_intermediate}, one recovers~\eqref{express_variance}.

\subsection{Proof of Result \ref{res2}}
If $s=-1$, then $z=0$ is a deterministic eigenvalue. This is natural, since
$\vb{I}+s\,u u^*$ is then non-invertible. Indeed,
\begin{equation}
    \vb{A}(\vb{I}-u u^*)u=0,
\end{equation}
so that $u$ is a right eigenvector associated with the eigenvalue $0$. In this case, the overlap is equal to $1$ and the fluctuations vanish.

Let us now look for the other possible outliers. By Eq.~\eqref{eq:finite_N_outlier_condition}, they must satisfy
\begin{equation}
    \bra{u}\vb{G}_{\vb{A}}(z)\vb{A}\ket{u}
    =
    -1,
\end{equation}
where
\begin{equation}
    \vb{G}_{\vb{A}}(z)
    =
    (z\vb{I}-\vb{A})^{-1}.
\end{equation}
Using the identity
\begin{equation}
    \vb{G}_{\vb{A}}(z)\vb{A}
    =
    z\vb{G}_{\vb{A}}(z)-\vb{I},
\end{equation}
this condition becomes
\begin{equation}
    z\,\bra{u}(z\vb{I}-\vb{A})^{-1}\ket{u}
    =
    0.
\end{equation}
The case $z=0$ has already been treated. Hence the nonzero outliers are given by the zeros of
\begin{equation}
    z\mapsto \bra{u}(z\vb{I}-\vb{A})^{-1}\ket{u}.
\end{equation}

Since $\vb{A}$ is rotationally invariant, one has, outside the spectral support,
\begin{equation}
    \bra{u}(z\vb{I}-\vb{A})^{-1}\ket{u}
    \longrightarrow
    \g_{\vb{A}}(z)
\end{equation}
in the high-dimensional limit. Therefore, if a sequence of nonzero outliers $(z_n)_n$ stays outside $\Omega_{\vb{A}}$ and converges to some point $z_*$, then $z_*$ must satisfy
\begin{equation}
    \g_{\vb{A}}(z_*)=0.
\end{equation}
In this regime, the overlap and the Gaussian fluctuations are then obtained from the same formulas as in Eqs.~\eqref{po} and~\eqref{eq:final_variance_intermediate}.

It remains to understand the nonzero outliers lying in $\Omega_{\vb{A}}$. Since $\g_{\vb{A}}$ vanishes identically on this domain, the deterministic equation $\g_{\vb{A}}(z)=0$ no longer selects isolated limiting locations. We are therefore led to study the zeros of the random analytic function
\begin{equation}
    z\in\Omega_{\vb{A}}
    \longmapsto
    \bra{u}(z\vb{I}-\vb{A})^{-1}\ket{u}.
\end{equation}
By rotational invariance, this has the same distribution as
\begin{equation}
    z\in\Omega_{\vb{A}}
    \longmapsto
    \left[(z\vb{I}-\vb{A})^{-1}\right]_{11}.
\end{equation}

We now explain why the Edelman--Kostlan mechanism naturally appears. For a large rotationally invariant matrix $\vb{M}$, set
\begin{equation}
    \vb{G}_{\vb{M}}(z)
    =
    (z\vb{I}-\vb{M})^{-1}.
\end{equation}
The fluctuation formula for diagonal entries gives, for a suitable matrix $\vb{B}$,
\begin{equation}
    [\vb{B}]_{11}
    =
    \tau(\vb{B})
    +
    \frac{1}{\sqrt N}\xi_{\vb{B}}
    +
    o(N^{-1/2}),
\end{equation}
where $\xi_{\vb{B}}$ is asymptotically a centered complex Gaussian variable with variance
\begin{equation}
    \mathbb{E}|\xi_{\vb{B}}|^2
    =
    \tau(\vb{B}\vb{B}^*)-|\tau(\vb{B})|^2.
\end{equation}
Applying this to the $z$-dependent matrix
\begin{equation}
    \vb{B}_z
    =
    \vb{G}_{\vb{M}}(z),
\end{equation}
we obtain
\begin{equation}
    [\vb{G}_{\vb{M}}(z)]_{11}
    =
    \tau(\vb{G}_{\vb{M}}(z))
    +
    \frac{1}{\sqrt N}f_N(z)
    +
    o(N^{-1/2}),
\end{equation}
where $f_N$ converges, in the sense of finite-dimensional distributions, to a centered Gaussian analytic function. Its pointwise variance is
\begin{equation}
    K_{\vb{M}}(z,z)
    =
    \tau\!\left(
    \vb{G}_{\vb{M}}(z)\vb{G}_{\vb{M}}(z)^*
    \right)
    -
    \left|
    \tau\!\left(\vb{G}_{\vb{M}}(z)\right)
    \right|^2.
\end{equation}
On $\Omega_{\vb{A}}$, the deterministic term vanishes, since
\begin{equation}
    \tau(\vb{G}_{\vb{A}}(z))=\g_{\vb{A}}(z)=0.
\end{equation}
Thus, on $\Omega_{\vb{A}}$, the variance reduces to
\begin{equation}
    K_{\vb{A}}(z,z)
    =
    h_{\vb{A}}(z),
\end{equation}
where
\begin{equation}
    h_{\vb{A}}(z)
    =
    \tau\!\left[
    \vb{G}_{\vb{A}}(z)\vb{G}_{\vb{A}}(z)^*
    \right]
    =
    \tau\!\left[
    \big((z\vb{I}-\vb{A})(z\vb{I}-\vb{A})^*\big)^{-1}
    \right].
\end{equation}

The zeros in $\Omega_{\vb{A}}$ are therefore governed, at leading order, by a Gaussian analytic function with variance profile $h_{\vb{A}}(z)$. Since these zeros correspond precisely to the nonzero outliers, the Edelman--Kostlan formula~\cite{EdelmanKostlan1995,Sodin2000} yields their limiting mean intensity:
\begin{equation}
    \varrho_{\vb{A}}^{\rm out}(z)
    =
    \frac{1}{4\pi}
    \Delta\log h_{\vb{A}}(z).
\end{equation}
Equivalently, for any domain $D\subset\Omega_{\vb{A}}$, denoting by
$N_{\vb{A}}^{\rm out}(D)$ the number of nonzero outliers lying in $D$, we have
\begin{equation}
    \mathbb{E}\,N_{\vb{A}}^{\rm out}(D)
    =
    \frac{1}{4\pi}
    \int_D
    \Delta\log h_{\vb{A}}(z)\,d^2z.
\end{equation}

Finally, if $z_N\in\Omega_{\vb{A}}$ is a nonzero outlier converging to $z$ and $\phi_N$ is its associated normalized right eigenvector, the eigenvector formula above, together with
\begin{equation}
    \langle u,(z_N\vb{I}-\vb{A})^{-1}u\rangle=0,
\end{equation}
gives
\begin{equation}
    |\langle u,\phi_N\rangle|^2
    \underset{N\to+\infty}{\longrightarrow}
    \frac{1}{1+|z|^2h_{\vb{A}}(z)}
    =
    1-|z|^2
    \partial_{\alpha}
    \mathcal{R}_{1,\vb{A}^{-1}}(0,z),
\end{equation}
which proves Eq.~\eqref{overbi}.

\bibliographystyle{plain}
\bibliography{References.bib}

\end{document}